# Comparative Analysis of Machine Learning Models for Short-Term Distribution System Load Forecasting

Elias Raffoul[1], Mingjian Tuo[2], Cunzhi Zhao[3], Tianxia Zhao[4], Meng Ling[4], and Xingpeng Li[1]

[1] Department of Electrical and Computer Engineering, University of Houston, Houston, TX, USA
[2] Key Laboratory of Automotive Power Train and Electronics, Hubei University of Automotive and Technology, Shiyan, China
[3] McNeese State University, Lake Charles, LA, USA
[4] Industrial AI Team, Shell, Houston, TX, USA

*Abstract*— Accurate electrical load forecasting is crucial for optimizing power system operations, planning, and management. As power systems become increasingly complex, traditional forecasting methods may fail to capture the intricate patterns and dependencies within load data. Machine learning (ML) techniques have emerged as powerful alternatives, offering superior prediction accuracy and the ability to model non-linear and complex temporal relationships. This study presents a comprehensive comparison of prominent ML models—feedforward neural networks, recurrent neural networks, long short-term memory networks, gated recurrent units, and the attention temporal graph convolutional network—for short-term load forecasting of the Energy Corridor distribution system in Houston, Texas. Using a 24-hour look-back window, we train the models on datasets spanning one and five years, to predict the load demand for the next hour and assess performance. Our findings aim to identify the most effective ML approach for accurate load forecasting, contributing to improved grid reliability and system optimization.

*Index Terms*— Attention Temporal Graph Convolutional Network, Distribution system, Energy Corridor, Gated Recurrent Unit, Load Forecasting, Long Short-Term Memory, Machine Learning, Recurrent Neural Network.

## I. INTRODUCTION

As energy consumption patterns evolve due to increasing electrification and renewable integration, the power grid faces unprecedented challenges in maintaining stability and reliability. Addressing these challenges requires advancements in predictive techniques that can anticipate demand fluctuations with greater precision. Modern approaches, particularly those leveraging machine learning (ML), have the potential to transform how power systems adapt to complex load behaviors.

Over the years, numerous approaches have been explored for load forecasting, ranging from classical time series models to more sophisticated statistical methods [1]. While these traditional techniques have provided valuable insights, their limitations in capturing the increasingly complex and dynamic behaviors of modern energy systems have become more apparent. The rise of ML has introduced a new era in forecasting, enabling models to learn from vast historical datasets and uncover intricate, non-linear patterns that were previously elusive [2]-[5].

This study conducts a comprehensive comparison of several notable ML models for short-term load forecasting of the Energy Corridor distribution system in Houston, Texas. The models evaluated include feedforward neural networks (FNN), recurrent neural networks (RNN), long short-term memory networks (LSTM), gated recurrent units (GRU), and the attention temporal graph convolutional network (A3T-GCN). Using a 24-hour look-back window, the models are trained on datasets spanning one year and five years, to predict the load demand for the following hour. Their performance is then assessed to evaluate their effectiveness.

FNNs have been successfully applied in various load forecasting studies, demonstrating their ability to model non-linear relationships [6]. However, they lack the capacity to effectively capture temporal dependencies. In contrast, RNN architectures have proven adept at learning from sequential data, with LSTM networks, in particular, showing considerable promise due to their ability to mitigate the vanishing gradient problem. Studies have demonstrated LSTM's superiority over standard RNNs and GRU models in load forecasting tasks, achieving the lowest mean absolute percentage error (MAPE) in their analysis of Panama City's load data [7].

LSTM's ability to learn long-term dependencies makes it a leading choice for load forecasting, and several studies have validated its effectiveness in capturing both short-term and long-term load trends. For example, a comparison of LSTM, GRU, and vanilla RNN models for load forecasting in the PJM East region found LSTM to consistently outperform its counterparts across multiple accuracy metrics, including root mean square error (RMSE) and mean absolute error (MAE) [8].

GRUs offer a simplified architecture compared to LSTMs while maintaining comparable performance [9]. Some research has demonstrated that GRUs can achieve equal or even superior results to LSTMs in specific load forecasting scenarios, benefiting from reduced computational complexity and faster training times [10].

Advancements in attention mechanisms have further enhanced the performance of sequence prediction models. The A3T-GCN integrates graph convolutional networks (GCN) with attention mechanisms to capture both spatial and temporal dependencies in time series data [11]. Originally developed for traffic forecasting, A3T-GCN shows promising potential for application in load forecasting due to its ability to model complex spatiotemporal relationships effectively.

While extensive research has been conducted on individual ML models for load forecasting, comparative analyses across different models and temporal scales are still relatively rare. Moreover, studies that focus on real-world distribution systems, particularly those that integrate both residential and commercial loads, are limited. This study addresses these gaps by offering a detailed comparison of FNN, RNN, LSTM, GRU, and A3T-GCN models, utilizing real-world datasets from the 44-bus Energy Corridor distribution system. Through this analysis, the study aims to uncover the strengths and weaknesses of each model in capturing short-term load patterns. The insights gained will not only enhance load forecasting accuracy but also inform the selection of the most appropriate models for various forecasting scenarios in power system operations.

## II. NETWORKS TOPOLOGIES

In this section, we describe the various network topologies of ML models employed in our study, highlighting the structure and unique characteristics of each.

### A. Feedforward Neural Networks (FNN)

FNNs are one of the most basic and widely used neural network architectures. As illustrated in Fig. 1, each neuron in a given layer is connected to every neuron in the subsequent layer, forming a dense network of connections. This architecture allows FNNs to capture complex patterns and relationships within the data by learning a series of weights that define the importance of each connection.

In the context of load forecasting, FNNs are advantageous due to their simplicity and ease of implementation. They can approximate any continuous function, making them versatile for various prediction tasks. However, FNNs do not inherently account for temporal or spatial dependencies within the data, which can limit their effectiveness in more complex scenarios, such as those involving time series or graph-structured data.

### B. Recurrent Neural Networks (RNN)

RNNs are a class of neural networks specifically designed for sequential data. Unlike FNNs, RNNs have connections that form directed cycles, allowing them to maintain a hidden state that evolves over time as they process sequences of data. This recurrent structure enables RNNs to model time-dependent phenomena by capturing the temporal dependencies within the data.

However, standard RNNs are prone to the vanishing gradient problem, which can make it difficult for them to learn long-term dependencies effectively [12]. Despite this limitation, RNNs are still useful for modeling short-term temporal dependencies in load predictions, providing insights into the immediate future behavior of the power system.

### C. Long Short-Term Memory (LSTM) Networks

LSTM networks are a type of RNN designed to address the vanishing gradient problem, which often hinders the training of traditional RNNs. LSTMs introduce a memory cell that maintains information over long periods, along with gates that control the flow of information into and out of the cell. This architecture allows LSTMs to capture long-term dependencies in sequential data, making them particularly well-suited for time series prediction tasks.

In load forecasting, LSTMs are used to model the temporal dynamics of the system, such as how load profiles or voltage levels evolve over time. Their ability to remember patterns across multiple time steps makes them effective for predicting future load behavior based on historical data. However, LSTMs do not explicitly consider the spatial relationships within the power grid, which can be a limitation when dealing with complex network structures.

### D. Gated Recurrent Units (GRU)

GRUs are a simplified version of LSTM networks that retain the ability to model long-term dependencies while being computationally less intensive. GRUs merge the forget and input gates of LSTMs into a single update gate, reducing the number of parameters and simplifying the training process. Despite their simpler structure, GRUs have demonstrated comparable performance to LSTMs in various sequence prediction tasks.

GRUs offer a balance between computational efficiency and predictive power for load estimation. They can effectively model the temporal aspects of the power system, capturing how different variables interact over time. Like LSTMs, GRUs focus on temporal dependencies and do not inherently capture spatial relationships within the grid. Fig. 1 compares FNN, RNN, LSTM, & GRU architectures.

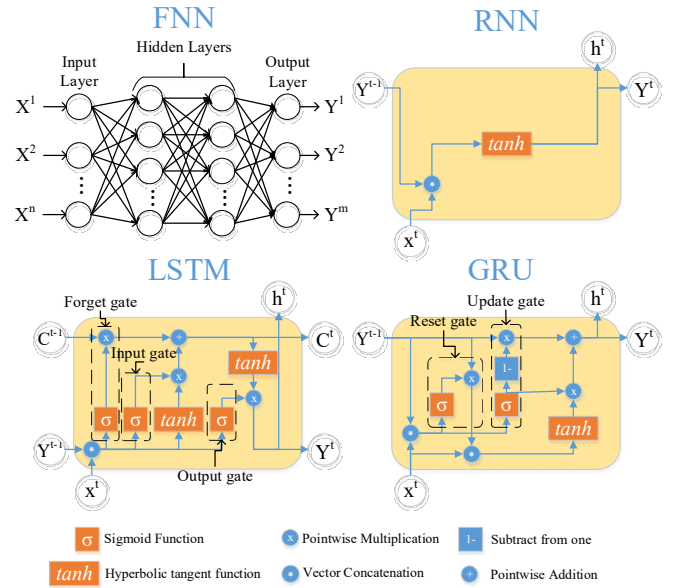

*Figure 1 – FNN, RNN, LSTM, & GRU Typical Block Architectures.*

### E. A3T-GCN Model

The A3T-GCN was designed with the dual purpose of simultaneously capturing global temporal dynamics and spatial correlations. This innovative model incorporates GRUs to grasp short-term trends in time series, and leverages GCNs to understand spatial dependencies based on the network's topology. Additionally, the A3T-GCN integrates an attention mechanism, allowing for the adjustment of the importance assigned to different time points seen in Fig. 2. This mechanism facilitates the aggregation of global temporal information, contributing to an enhancement in prediction accuracy [11]. Notably, Fig. 2 reveals an intriguing parameter in this model, allowing us to specify the number of time periods for which we aim to make predictions into the future— an option that adds flexibility and adaptability to the A3T-GCN.

GCNs operate on graph-structured data, making them well-suited for power systems, which can be naturally represented as graphs where buses are nodes and transmission lines are edges. They use convolutional operations to process information from neighboring nodes, capturing spatial dependencies within the network. By combining GRU with GCN, the A3T-GCN can model both the temporal changes of power system variables and the spatial interactions between grid components. This dual capability allows for more accurate load predictions, especially in scenarios where both time and network structure are important.

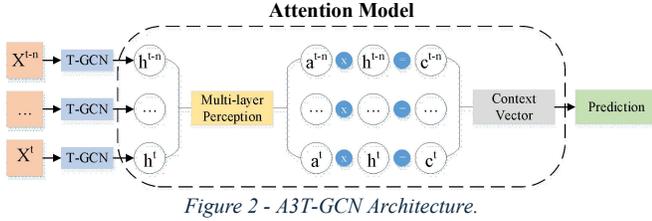

*Figure 2 - A3T-GCN Architecture.*

### F. Summary

Each of these network topologies brings unique strengths to load prediction analysis. FNNs offer simplicity and broad applicability, while LSTMs and GRUs excel at modeling temporal dependencies. RNNs provide a more general approach to sequence modeling, though they are less effective for long-term dependencies. The A3T-GCN model stands out by incorporating both temporal and spatial information, making it particularly well-suited for the complex, interconnected nature of electricity distribution networks. Understanding these topologies is essential for selecting the most appropriate ML model for a given load forecasting task.

## III. SYNTHETIC ENERGY CORRIDOR DISTRIBUTION NETWORK: SYSTEM CREATION & DATASET GENERATION

Fig. 3 showcases the load nodes placement in the Energy Corridor. The synthetic system comprises 44 load nodes and 42 three-phase distribution lines.

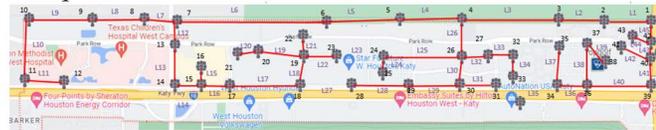

*Figure 3 - Nodal loads in Energy Corridor Distribution System.*

As depicted in Fig. 4, the area is geographically divided by roads and communities. To accurately represent power consumption within each nodal area, it is essential to consider the prevailing load type in the segmented region. By leveraging the size of the region and average consumption data, nodal load values can be estimated. In the Energy Corridor area, two main types of loads exist: domestic (residential) and commercial loads. Domestic load encompasses the aggregate energy consumed by household electrical appliances, including lights, refrigerators, heaters, and air conditioners. As per the Environmental Protection Agency and Energy Information Administration, the average daily electricity consumption is recorded at 37 kWh [13]. On the other hand, the Department of Energy (DOE) suggests an average of approximately 22.5 kWh per square foot per year for commercial buildings [14].

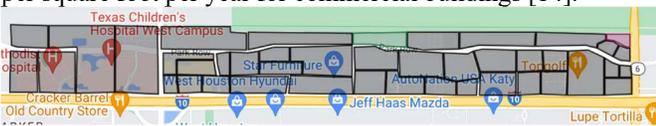

*Figure 4 - Nodal Areas in Energy Corridor Distribution System.*

The residential base load profiles are developed using EnergyPlus, a comprehensive building energy simulation program that models energy consumption for heating, cooling, lighting, and other building systems. In this study, five EnergyPlus models represent single-family detached homes built according to the 2009 International Energy Conservation Code (IECC) standards. These models are tailored to different climate regions and simulated across various Typical Meteorological Year (TMY3) locations to accurately reflect conditions in the Houston area.

In contrast, commercial load profiles are based on 16 ASHRAE 90.1-2004 DOE Commercial Prototype Models, which are also simulated using TMY3 data [15]. These models adjust insulation levels according to ASHRAE 90.1-2004 requirements for each climate zone. The distribution system includes five types of commercial loads: hospitals, restaurants, retail stores, hotels, and offices. To create a comprehensive five-year load profile, the base load values are scaled according to Houston's specific percentage profiles, with the generated profile data recorded at hourly intervals.

To organize the dataset effectively, it's crucial to consider the structure of our distribution power system, comprising 44 nodes buses. The input dataset for the models is formatted as an array of dimensions 8760 (1 year) or 43800 (5 years) by 88, where each entry represents the nodal bus reactive (Q) and active (P) loads for every hour of the year. In specific terms, the input data encompasses the P and Q values for each of the 44 buses, spanning a defined look-back period and batch size. This is represented as [Batch Size, Look-back, 88]. For instance, with a batch size of 1 and a look-back of 24 hours, the input data would be structured as [1, 24, 88]. In the context of predicting the $25^{th}$ bus load (P, Q), considering a batch size of 1 and a look-back of 24 hours, the output data would be a matrix of dimensions (1, 88).

## IV. RESULTS

Following training over 100 epochs, each model was evaluated using MAE, MSE, and MAPE metrics. The performance results are summarized in Table 1.

TABLE I - COMPARISONS OF ML MODELS TRAINED ON 1 & 5 YEARS DATASETS

| Dataset Size | Model | MAE | MSE | MAPE |
|---|---|---|---|---|
| 1 year | FNN | 26.94 | 2510.29 | 32.35 |
| | LSTM | 37.09 | 5268.33 | 29.95 |
| | GRU | 32.01 | 3715.34 | 26.09 |
| | RNN | 27.16 | 2711.71 | 25.58 |
| | A3T-GCN | 41.10 | 6848.65 | 85.28 |
| 5 years | FNN | 4.82 | 84.84 | 25.21 |
| | LSTM | 5.80 | 129.41 | 23.59 |
| | GRU | 5.46 | 111.52 | 24.57 |
| | RNN | 4.70 | 80.47 | 21.05 |
| | A3T-GCN | 8.09 | 253.12 | 84.44 |

For the one-year dataset, RNN achieves the lowest MAPE of 25.58, indicating it has the smallest percentage error among the models, making it highly effective in capturing relative forecast accuracy. However, FNN shows better performance in terms of MAE and MSE, with lower values suggesting it makes fewer absolute and squared errors compared to the other models. GRU also performs well, showing better results than LSTM. A3T-GCN, on the other hand, has higher error metrics across all measures, suggesting it is less effective for short-term load forecasting.

In the five-year dataset, all models show improved accuracy compared to the one-year dataset. RNN exhibits the strongest performance with the lowest MAE, MSE, and MAPE, closely followed by FNN. LSTM and GRU also demonstrate enhanced performance, with GRU providing competitive results. A3T-GCN remains the least effective model for long-term forecasting, similar to its performance on the shorter dataset.

Fig. 5 illustrates the active and reactive loads for bus 14, showing both the ground truth and the predicted values of each model trained on a 5-year dataset. Most models closely align with the actual values, demonstrating strong predictive performance.

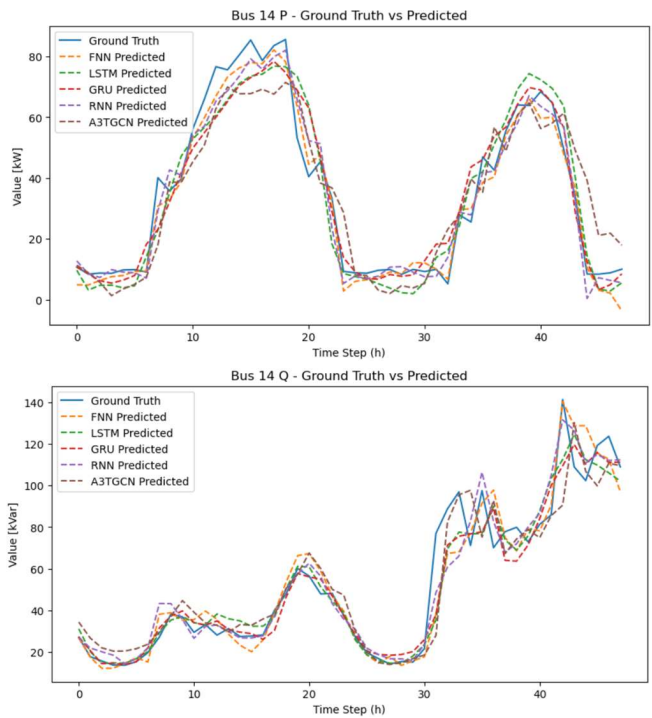

*Figure 5 - Active (P) & Reactive (Q), Exact & Predicted Loads for Bus 14 Using Models Trained on a 5-Year Dataset*

In summary, RNN & FNN demonstrate the highest accuracy in short-term forecasting, sharing the lowest error values across datasets. GRU and LSTM also show strong performance, particularly with longer datasets, though LSTM surpasses GRU in MAPE. FNN performs well in MAE and MSE but is less effective in relative accuracy compared to RNN. Despite its advanced architecture, A3T-GCN performs the least effectively across all metrics

Fig. 6 illustrates the accuracy of each model at various error tolerance levels (10%, 15%, and 20%). For the one-year dataset, accuracy generally improves as the error tolerance increases across all models. The FNN consistently delivers the highest accuracy at all tolerance levels, while the RNN model usually ranks second. FNN achieves 30% of its values within a 10% error margin, 40% within a 15% error margin, and 50% within a 20% error margin. The A3T-GCN consistently shows the lowest accuracy across all scenarios. Most models exhibit significant accuracy gains during the first 20-30 epochs, after which the rate of improvement slows considerably.

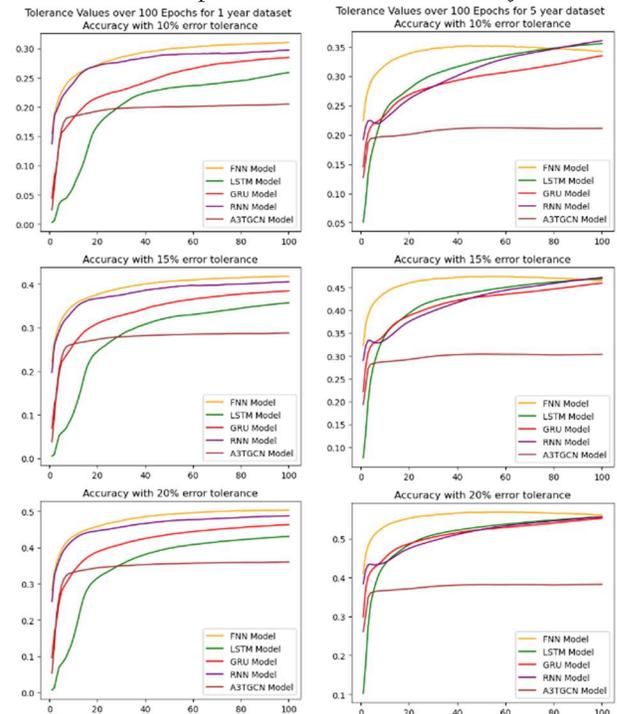

*Figure 6 – Accuracy Metrics Under Various Tolerance Conditions.*

In the five-year dataset, FNN significantly outperforms other models, particularly during the initial epochs, though it begins to underperform relative to LSTM and RNN around the 60th epoch for the 10% &15% error tolerance. The FNN model achieved a 5% increase in its values for each error margin compared to the 1-year data. A3T-GCN still has the lowest accuracy, with the performance gap with other models being wider compared to the one-year dataset. LSTM, GRU, and RNN perform comparably, with minor variations depending on error tolerance. All models, except A3T-GCN, show rapid accuracy improvement in the first few epochs, with performance leveling off more quickly than observed in the one-year dataset.

Overall, accuracy levels are higher in the five-year dataset, particularly at higher error tolerances. The performance disparity between models (excluding A3T-GCN) is reduced in the five-year dataset, indicating that the additional data may enhance learning efficiency and accelerate peak performance achievement.

The observed superior performance of simpler models such as FNN and RNN compared to more complex models like

A3T-GCN can be attributed to several factors. Complex models like A3T-GCN are designed to capture detailed spatiotemporal patterns by combining GCNs with attention mechanisms. While these models are powerful, they can suffer from overfitting if the training data isn't diverse enough. Overfitting happens when a model starts to learn the noise in the data rather than the actual patterns, which reduces its performance on new data.

Simpler models like FNNs and RNNs, with fewer parameters, are less prone to overfitting. Their simplicity makes them more robust to variations and noise in the data, so they often perform well even with smaller or less diverse datasets. These models are generally better at identifying key patterns without the added complexity that can make more advanced models unstable or less generalizable.

Training requirements also differ. Complex models like A3T-GCN need a lot of data and fine-tuning to reach their full potential. If these conditions aren't met, their advanced features might not be utilized effectively. In contrast, simpler models can perform well with less data and require less computational power and training time, allowing for more frequent adjustments during development. The computational efficiency of simpler models also contributes to their effectiveness. FNNs and RNNs are less resource-intensive and quicker to train (about 40 seconds per epoch) compared to complex models like A3T-GCN, which can take up to 16 minutes per epoch. This efficiency helps in iterative model development and tuning.

Lastly, how models handle feature extraction and interpretability matters too. FNNs focus on straightforward relationships in the data, which is beneficial for simpler patterns. A3T-GCN aims to capture complex relationships but can sometimes introduce noise or obscure patterns, especially when the dataset is simpler.

## V. CONCLUSION

In conclusion, our analysis of ML models for short-term load forecasting in the Energy Corridor distribution system highlights some key findings. The RNN and FNN models emerged as the top performers. The GRU and LSTM models also showed strong results, especially with larger datasets, with GRU often outperforming LSTM in terms of MAE and MSE.

However, advanced models like A3T-GCN, while capable of capturing complex relationships, faced challenges such as overfitting and high computational demands, which affected their performance. On the other hand, simpler models like FNNs and RNNs proved to be effective, especially when data is limited or less complex.

It is important to note that these results are specific to the Energy Corridor distribution system and might vary for different systems or datasets. Future research could focus on improving advanced models like A3T-GCN through better tuning and additional features. Additionally, while the A3T-GCN model may underperform in short-term forecasting scenarios due to overfitting and its need for large, diverse datasets, it may excel in long-term forecasting tasks. The ability of A3T-GCN to capture intricate spatiotemporal relationships through GCNs and attention mechanisms could be more advantageous when modeling extended time horizons, where complex interactions between variables play a more significant role. Further research and experimentation with long-term datasets could reveal the full potential of A3T-GCN.

ACKNOWLEDGMENT

This work was financially supported by Shell. The authors acknowledge Jason Kowalski for his inputs for this research.